\documentclass[twocolumn,floatfix,altaffilletter,superscriptaddress,preprintnumbers,
               tightenlines,showpacs,showkeys]{revtex4-1}
\usepackage[utf8]{inputenc}
\usepackage[colorlinks=true,citecolor=blue,linkcolor=blue]{hyperref}
\usepackage[normalem]{ulem}
\usepackage{amsmath,amssymb}
\usepackage{epsfig}
\usepackage{graphicx}               
\usepackage{url}
\usepackage{color}
\usepackage{slashed}
\usepackage{multirow}
\usepackage{placeins}
\usepackage[dvipsnames]{xcolor}
\usepackage{epstopdf}
\usepackage{soul}
\usepackage{tikz}
\usetikzlibrary{trees}
\usetikzlibrary{decorations.pathmorphing}
\usetikzlibrary{decorations.markings}

\definecolor{jblue}  {RGB}{20,50,100}
\definecolor{npurple}  {RGB} {153, 51, 204}
\definecolor{wred}   {RGB}{217,0,56}
\definecolor{white}   {RGB}{255,255,255}

\definecolor{korange}   {RGB}{235, 80,  43}
\definecolor{korange2}   {RGB}{245, 100,  63}
\definecolor{kyelloworange}   {RGB}{255, 210,  110}
\definecolor{kyelloworange2}   {RGB}{240, 170,  90}
\definecolor{kred}   {RGB}{204,  102, 153}
\definecolor{kpurple}   {RGB}{153,  61, 190}
\definecolor{kpurplelight}   {RGB}{213,  161, 230}

\usepackage{cleveref}
\usepackage{subfigure} 
\usepackage{lipsum}
\definecolor{red}{rgb}{1.0, 0, 0}
 \usepackage{gensymb}
\allowdisplaybreaks

\bibpunct{[}{]}{,}{n}{}{,}

\newcommand{\ev}[1]{\ensuremath{\left\langle #1 %
                     \right\rangle}} 

\newcommand{\BR}{\text{BR}}


\begin{document}

\title{X-Ray Lines from DM Annihilation at the keV Scale}

\author{Vedran Brdar\footnote{vbrdar@uni-mainz.de }}
\affiliation{PRISMA Cluster of Excellence and
             Mainz Institute for Theoretical Physics,
             Johannes Gutenberg-Universit\"{a}t Mainz, 55099 Mainz, Germany}
\affiliation{Max-Planck-Institut f\"ur Kernphysik, Saupfercheckweg 1,
             69117 Heidelberg, Germany}

\author{Joachim Kopp\footnote{jkopp@uni-mainz.de}}
\affiliation{PRISMA Cluster of Excellence and
             Mainz Institute for Theoretical Physics,
             Johannes Gutenberg-Universit\"{a}t Mainz, 55099 Mainz, Germany}

\author{Jia Liu\footnote{liuj@uni-mainz.de}}
\affiliation{PRISMA Cluster of Excellence and
             Mainz Institute for Theoretical Physics,
             Johannes Gutenberg-Universit\"{a}t Mainz, 55099 Mainz, Germany}
\affiliation{Enrico Fermi Institute, University of Chicago, Chicago, IL 60637, USA}

\author{Xiao-Ping Wang\footnote{xiaowang@uni-mainz.de}}
\affiliation{PRISMA Cluster of Excellence and
             Mainz Institute for Theoretical Physics,
             Johannes Gutenberg-Universit\"{a}t Mainz, 55099 Mainz, Germany}
\affiliation{High Energy Physics Division, Argonne National Laboratory,
             Argonne, IL 60439, USA}

\date{\today}

\preprint{MITP/17-060 ~ EFI-17-22}

\begin{abstract}
  In 2014, several groups have reported hints for a yet unidentified line in
  astrophysical X-ray signals from galaxies and galaxy clusters at an energy
  of 3.5\,keV.  While it is not
  unlikely that this line is simply a reflection of imperfectly modeled atomic
  transitions, it has renewed the community's interest in models of keV-scale
  dark matter, whose decay would lead to such a line.  The alternative
  possibility of dark matter annihilation into monochromatic photons is far
  less explored, a lapse that we strive to amend in this paper. More precisely,
  we introduce a novel model of fermionic dark matter $\chi$ with
  $\mathcal{O}(\text{keV})$ mass, annihilating to a scalar state $\phi$ which
  in turn decays to photons, for instance via loops of heavy vector-like
  fermions.  The resulting photon spectrum is box-shaped, but if $\chi$ and
  $\phi$ are nearly degenerate in mass, it can also resemble a narrow line.  We
  discuss dark matter production via two different mechanisms -- misalignment
  and freeze-in -- which both turn out to be viable in vast regions of
  parameter space.  We constrain the model using astrophysical X-ray data, and
  we demonstrate that, thanks to the velocity-dependence of the annihilation
  cross section, it has the potential to reconcile the various observations of
  the 3.5\,keV line.  We finally argue that the model can easily avoid
  structure formation constraints on keV-scale dark matter.
\end{abstract}

\maketitle


Many recent papers in astroparticle physics start out by exposing the waning
of traditional dark matter (DM) candidates, especially the Weakly Interacting Massive
Particle (WIMP)~\cite{Arcadi:2017kky,Duerr:2016tmh, Bernal:2017kxu}. 
The present article is no exception.
And while it is certainly too early to give up on the elegance
of the thermal freeze-out mechanism for DM production, a look beyond is now
more motivated than ever.

In this paper, we will dwell on the possibility that DM is a fermion with a
mass of only a few keV. Such DM candidates, which often go by the name of ``sterile
neutrinos'' are well known for their potential to improve predictions for small
scale structure (see for instance ref.~\cite{Abazajian:2009hx} and references
therein), their clean observational signatures in the form
of X-ray lines \cite{Horiuchi:2013noa, Boyarsky:2005us, Abazajian:2006jc,
  Abazajian:2009hx, Bulbul:2014sua, Sekiya:2015jsa},
  and their tendency to evoke animated discussions
among physicists \cite{Jeltema:2014qfa, Boyarsky:2014paa, Bulbul:2014ala,
Jeltema:2014mla, Cline:2014vsa}. Many such discussions were
incited by recent claims for a yet-unidentified line at $\sim 3.5$\,keV
in the X-ray spectra from galaxies and galaxy clusters
\cite{
  Bulbul:2014sua,    
  Boyarsky:2014jta,  
  Riemer-Sorensen:2014yda, 
  Jeltema:2014qfa,   
  Boyarsky:2014paa,  
  Boyarsky:2014ska,  
  Malyshev:2014xqa,  
  Bulbul:2014ala,    
  Lovell:2014lea,    
  Jeltema:2014mla,   
  Carlson:2014lla,   
  Tamura:2014mta,    
  Iakubovskyi:2015dna, 
  Iakubovskyi:2015kwa, 
  Iakubovskyi:2015wma, 
  Savchenko:2015eiq, 
  Jeltema:2015mee,   
  Ruchayskiy:2015onc, 
  Chan:2016aab,      
  Franse:2016dln,    
  Bulbul:2016yop,    
  Hofmann:2016urz,   
  Conlon:2016lxl,    
Shah:2016efh}.      
Whether the origin of this lines is
indeed related to DM physics
\cite{
  Abazajian:2001vt,
  Ko:2014xda,        
  Queiroz:2014yna,   
  Babu:2014pxa,      
  Bezrukov:2014nza,  
  Lee:2014koa,       
  Chakraborty:2014tma, 
  Ishida:2014fra,    
  Geng:2014zqa,      
  Haba:2014taa,      
  Cline:2014kaa,     
  Hamaguchi:2014sea, 
  Nakayama:2014cza,  
  Alvarez:2014gua,   
  Kang:2014cia,      
  Biswas:2015sva,    
  Berlin:2015sia,    
  Roland:2015yoa,    
  Borah:2016ees,     
  Ishida:2014dlp,    
  Finkbeiner:2014sja, 
  Higaki:2014zua,    
  Jaeckel:2014qea,   
  Krall:2014dba,     
  Nakayama:2014ova,  
  Kong:2014gea,      
  Frandsen:2014lfa,  
  Choi:2014tva,      
  Cicoli:2014bfa,    
  Aisati:2014nda,    
  Lee:2014xua,       
  Kolda:2014ppa,     
  Bomark:2014yja,    
  Liew:2014gia,      
  Allahverdi:2014dqa, 
  Demidov:2014hka,   
  Dudas:2014ixa,     
  Cline:2014eaa,     
  Conlon:2014xsa,    
  Conlon:2014wna,    
  Chiang:2014xra,    
  Falkowski:2014sma, 
  Cline:2014vsa,     
  Cheung:2014tha,    
  Agrawal:2015tfa,   
  Lee:2015sha,       
  Heikinheimo:2016yds, 
  Arguelles:2016uwb, 
  Cosme:2017cxk,     
  Heeck:2017xbu,     
  Bae:2017tqn,       
  Boulebnane:2017fxw}, 
or simply to imperfect modeling of atomic physics effects
\cite{Jeltema:2014qfa, Jeltema:2014mla, Shah:2016efh, Aharonian:2016gzq},
the controversy surrounding it
cannot belie the fact that precision observations of the X-ray sky are a
prime tool to search for keV-scale DM.  While the overwhelming majority of studies
on this topic focus on line signals from DM \emph{decay}, we will in the following
explore the possibility that such signals arise from DM \emph{annihilation}.

\begin{figure}
  \centering
  \begin{tabular}{cc}
    \multicolumn{2}{l}{\includegraphics[width=0.9\columnwidth]{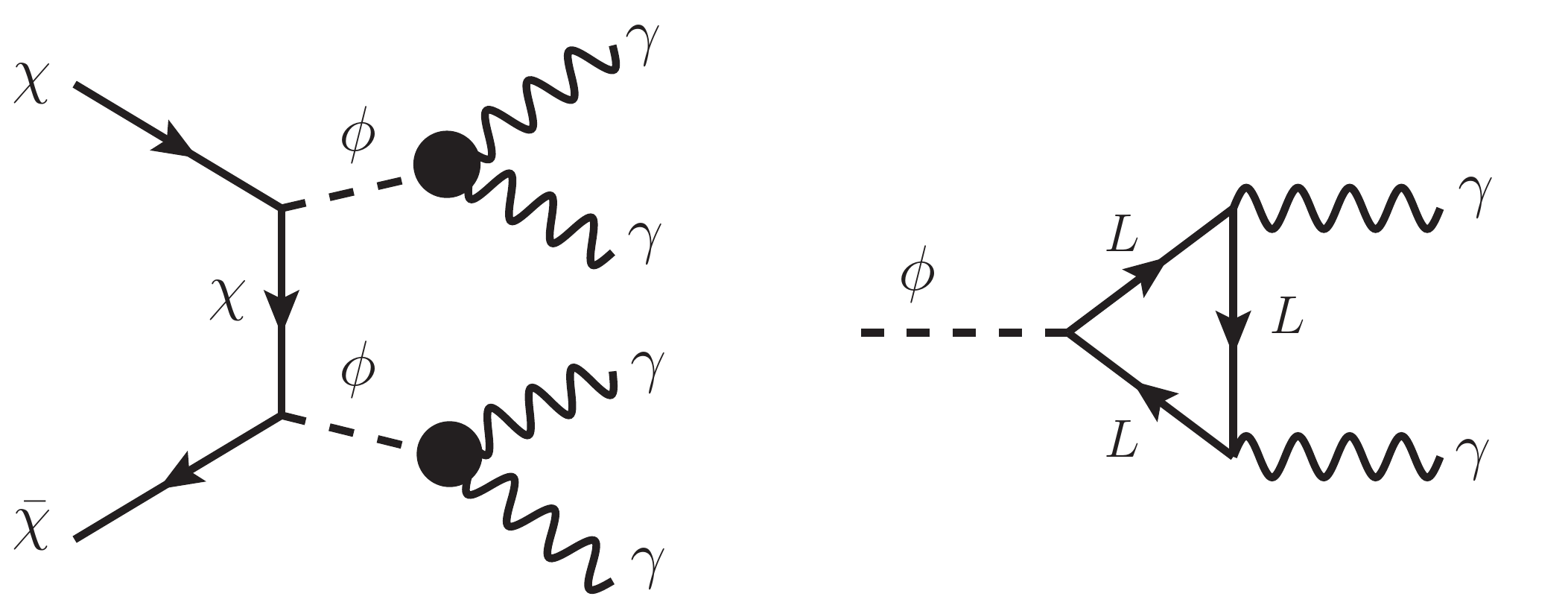}} \\
    \hspace{1cm} (a) & \hspace{2cm} (b)
  \end{tabular}
  \caption{(a) Feynman diagram for DM annihilation to photons via
    the intermediate scalar $\phi$. Blobs represent the effective $\phi \gamma
    \gamma$ coupling from \cref{eq:L-eff}. Diagram (b) shows a possible
    UV completion for this vertex, see \cref{eq:Lagrangian}.}
  \label{fig:feynman}
\end{figure}

We will introduce a toy model in which keV-scale Dirac fermions
$\chi$ annihilate to pairs of new real scalars $\phi$, which in turn
decay to photons, see \cref{fig:feynman} (a).  These photons will have a
box-shaped spectrum, which can be indistinguishable from a monochromatic line
within experimental resolutions if $\chi$ and $\phi$ are nearly degenerate in
mass.  We will see below that the model can also explain the
observed DM relic abundance.
At low energies, the phenomenology of our model is
captured by an effective Lagrangian consisting of just the $\phi$ couplings to
photons and $\chi$:
\begin{align}
  \mathcal{L}_\text{eff} \supset
      \tfrac{\alpha}{4\pi \Lambda} F_{\mu\nu} F^{\mu\nu} \phi
    + y \phi \bar\chi \chi \,,
  \label{eq:L-eff}
\end{align}
where $y$ is a new dimensionless coupling constant, $F^{\mu\nu}$ is the
electromagnetic field strength tensor, $\alpha$ is the electromagnetic fine
structure constant, and $\Lambda$ is the suppression scale of the dimension-5
interaction. We will see below that explaining the 3.5\,keV line while
satisfying all constraints requires $y \sim \text{few} \times 10^{-5}$ and
$\Lambda \sim 10$--$100$\,PeV.

\Cref{eq:L-eff} can be completed to a
renormalizable model for instance by introducing heavy vector-like leptons $L$
with charge assignment $(0,1,-2)$ under $SU(3)_c\times SU(2)_L\times U(1)_Y$:
\begin{align}
  \mathcal{L} &\supset
    \, \bar{L} i\slashed{D}\,L
     - m_L \bar{L} L
     + y \phi \bar\chi \chi
     + g \phi \bar{L} L \,,
  \label{eq:Lagrangian}
\end{align}
where $D^\mu$ is the gauge covariant derivative, $m_L$ is the mass of the
heavy leptons, and $g$ is the dimensionless $L$--$\phi$ coupling constant.  
In this UV completion,
\begin{align}
  \frac{1}{\Lambda} = \frac{4 g m_L}{\mu^2} f(4 m_L^2/ \mu^2)  \,,
  \label{eq:c-over-Lambda}
\end{align}
where $\mu$ is the renormalization scale, which we take equal to $m_\phi$,
and the loop function corresponding to the diagram in \cref{fig:feynman}~(b)
is given by \cite{Marciano:2011gm}
\begin{align}
  f(\tau) = \left\{ \begin{array}{ll}
                      1 - (\tau -1)(\csc^{-1}\sqrt{\tau})^2) &  \tau \geqslant 1 \\
                      1 + \frac{\tau-1}{4} \left[\log[\frac{1+\sqrt{1-\tau}}
                                                           {1-\sqrt{1-\tau}}] -i\pi\right]^2
                                                             &  \tau < 1
                    \end{array} \right. \,.
  \label{eq:f}
\end{align}

\vspace{1ex}
{\bf DM Annihilation.}
The cross section for the DM annihilation process $\bar\chi \chi\to \phi\phi$
depends on the relative mass difference
$\delta \equiv (m_{\chi} - m_\phi)/m_{\chi}$ of $\chi$ and $\phi$
and on the relative velocity $v_\text{rel}$ of the annihilating
DM particles. In the Milky Way, $v_\text{rel} \sim 200\,\text{km/sec}$, while
in galaxy clusters, $v_\text{rel} \sim 1\,000\,\text{km/sec}$.
If $|\delta| \ll v_\text{rel}^2$, the annihilation cross section is
\begin{align}
  \sigma_\text{ann} v_\text{rel}
    = \frac{y^4 v_\text{rel}^3}{16\pi m_{\chi}^2} \,.
  \label{eq:sigmav1}
\end{align}
For $\delta \gg v_\text{rel}^2$, we find
\begin{align}
  \sigma_\text{ann} v_\text{rel}
    = \frac{\sqrt{\delta} (2 \delta (\delta+2) + 3) \, y^4 v_\text{rel}^2}
           {24 \pi (1+\delta)^4 \, m_{\chi}^2} \,.
  \label{eq:sigmav2}
\end{align}
Note the different dependence on $v_\text{rel}$ in these two limiting cases.  A
small value of $\delta$ could be naturally explained in a supersymmetric
extension of the model, with $\chi$ and $\phi$ residing in the same
supermultiplet~\cite{ArkaniHamed:2008qp, Baumgart:2009tn, Cheung:2009qd}.

The decay rate of $\phi$ to two photons is $\Gamma_{\phi \to \gamma\gamma} =
\alpha^2 m_\phi^3 / [(4\pi)^3 \Lambda^2]$.  For DM annihilation in the Milky
Way, this implies that the morphology of the photon signal traces the
distribution of DM only if $\Lambda \lesssim 2.5\,\text{PeV} \times (m_\phi /
\text{keV})^3$. Then, the intermediate $\phi$ particles travel for $\lesssim
1$\,kpc before decaying, so the wash-out of the annihilation signal due to
their long lifetime is degenerate with the large uncertainties in the Milky
Way's DM halo profile at radius $\lesssim 1$\,kpc~\cite{Chu:2017vao}.

\begin{figure}
  \centering
  \includegraphics[width=0.95\columnwidth]{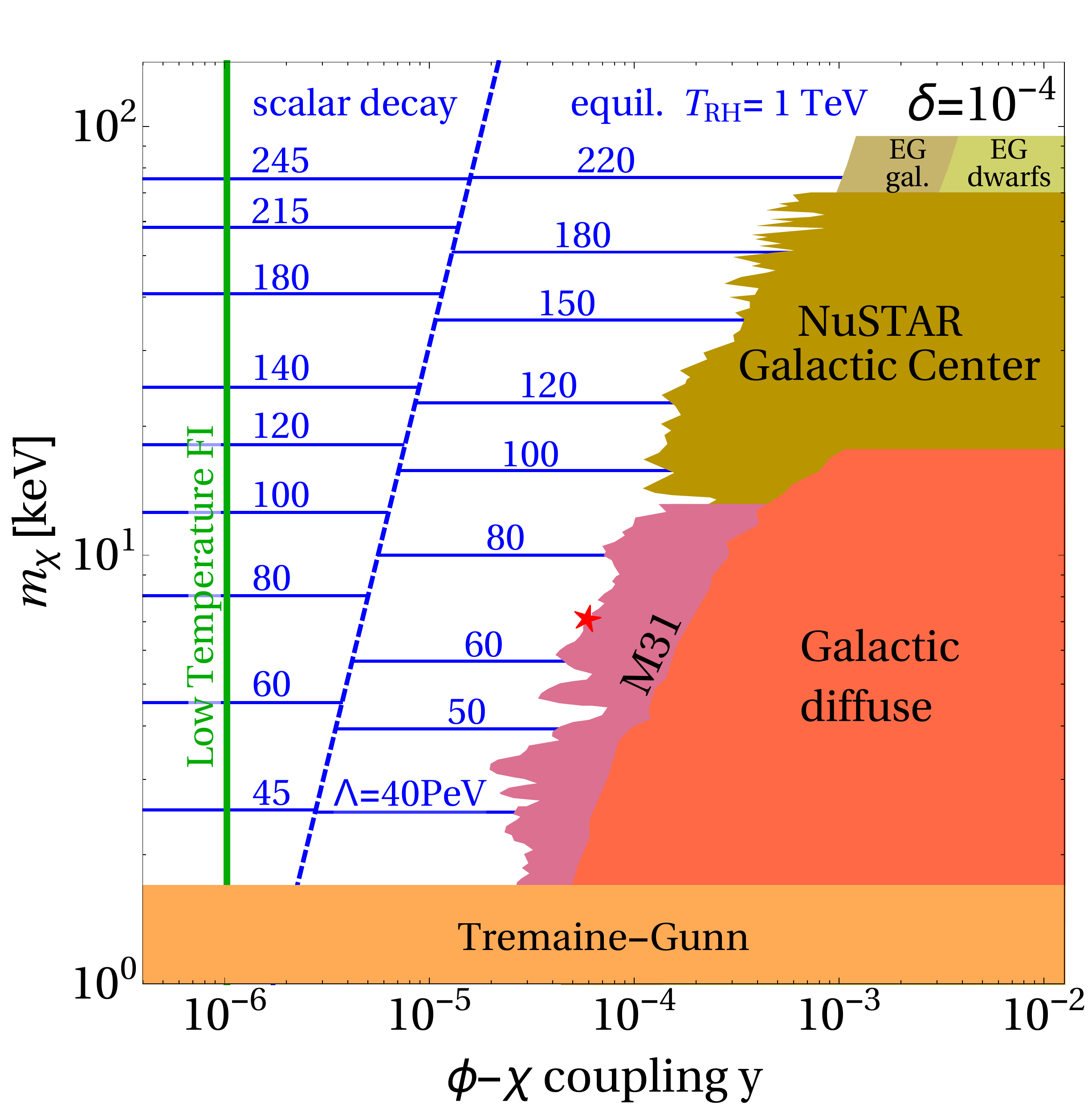}
  \caption{Parameter space of the annihilating keV-scale DM models defined
    in \cref{eq:L-eff,eq:Lagrangian}.  We show the relevant constraints as a
    function of the DM mass $m_{\chi}$ and the Yukawa coupling between $\chi$
    and the mediator $\phi$, assuming a degeneracy parameter of $\delta \equiv
    (m_{\chi} - m_\phi) / m_{\chi} = 10^{-4}$.  Shaded regions are excluded by
    X-ray observations of the Andromeda Galaxy (M31) \cite{Horiuchi:2013noa},
    the Galactic Center (``NuSTAR Galactic Center'') \cite{Perez:2016tcq}, the
    galactic (``Galactic diffuse'') \cite{Abazajian:2006jc} and extragalactic
    (``EG.~gal'' and ``EG.~dwarfs'') \cite{Boyarsky:2005us} diffuse X-ray
    background, and by Pauli blocking arguments (``Tremaine--Gunn'')
    \cite{Tremaine:1979we, Iakubovskyi:2013}.  The red star~\cite{redstar}
    corresponds to the excess of monochromatic X-rays at 3.5\,keV observed in
    ref.~\cite{Bulbul:2014sua, Boyarsky:2014jta}.  Horizontal blue lines
    indicate the suppression scale $\Lambda$ of the $\phi\gamma\gamma$ coupling
    required to obtain the correct DM relic abundance via UV freeze-in,
    assuming $T_\text{RH} = 1$\,TeV. To the right of the dashed line,
    part of the initially produced abundance of $\chi$ gets transferred to $\phi$ via
    $\phi\phi \leftrightarrow \bar\chi \chi$ at $T \sim \text{keV}$.  The
    region to the right of the dashed line can also be reached via freeze-in through
    misalignment (see text).  The vertical green band indicates where low temperature
    freeze-in via $\phi\phi \to \bar\chi\chi$ yields the correct relic abundance.}
  \label{fig:paramspace}
\end{figure}

For $\delta \lesssim 10^{-5}$--$10^{-4}$ so that the
photon signal from DM annihilation appears monochromatic within experimental
energy resolutions, we compare the signals predicted by our model to data in
\cref{fig:paramspace}.
The shaded exclusion regions in this plot are based on reinterpreting existing
limits on DM \emph{decay} to monochromatic photons by equating the photon flux
in the two cases:
\begin{align}
  &\frac{\Gamma_{\chi}}{4\pi m_{\chi}}
   \int \! dl \,d\Omega \, \rho_\text{DM}(l,\Omega)  =
                                                        \nonumber\\
  &\quad
   \frac{4}{16\pi m_{\chi}^2}
   \int \! dl\, d\Omega \, d^3 v_\text{rel} \,
           \rho^2_\text{DM}(l,\Omega) \,\sigma_\text{ann} v_\text{rel} \,
           f(l, \Omega, v_\text{rel}) \,.
   \label{eq:dec-ann-flux}
\end{align}
Here, $\Gamma_{\chi}$ is the DM decay rate,
$\rho_\text{DM}$ is the DM density, and $f(l, \Omega, v_\text{rel})$ is the DM
velocity distribution at distance $l$ along the line of sight in solid angle
direction $\Omega$.  We obtain $f(l, \Omega, v_\text{rel})$ numerically by
evaluating Eddington's formula~\cite{Lisanti:2010qx} (see \cite{Ferrer:2013cla}
for possible shortcomings of this approach due to baryonic effects).
The factor 4 on the right hand side of \cref{eq:dec-ann-flux} accounts for the
number of photons produced in each annihilation $\bar\chi\chi \to \phi\phi \to
4\gamma$.  For the limits based on diffuse extragalactic X-rays from galaxies
and dwarf galaxies (``EG gal.'' and ``EG dwarfs''), \cref{eq:dec-ann-flux}
receives redshift-dependent corrections~\cite{Hooper:2007be,Kopp:2015bfa}.
Among these is a factor $\Delta^2(z) \equiv \Delta^2(0)/(1+z)^3$
\cite{Bergstrom:2001jj,Hooper:2007be} that accounts for the stronger clumping
of DM at late times in cosmological history.  We conservatively choose
$\Delta^2(0) = 10^6$ \cite{Taylor:2002zd}.  We moreover assume that the DM
velocity distributions $f(l,\Omega,v_\text{rel})$ of distant galaxies and dwarf
galaxies follow those of the Milky Way and of Milky Way dwarfs, respectively.

\begin{figure}
  \centering
  \begin{tabular}{c}
    \includegraphics[width=0.4\textwidth]{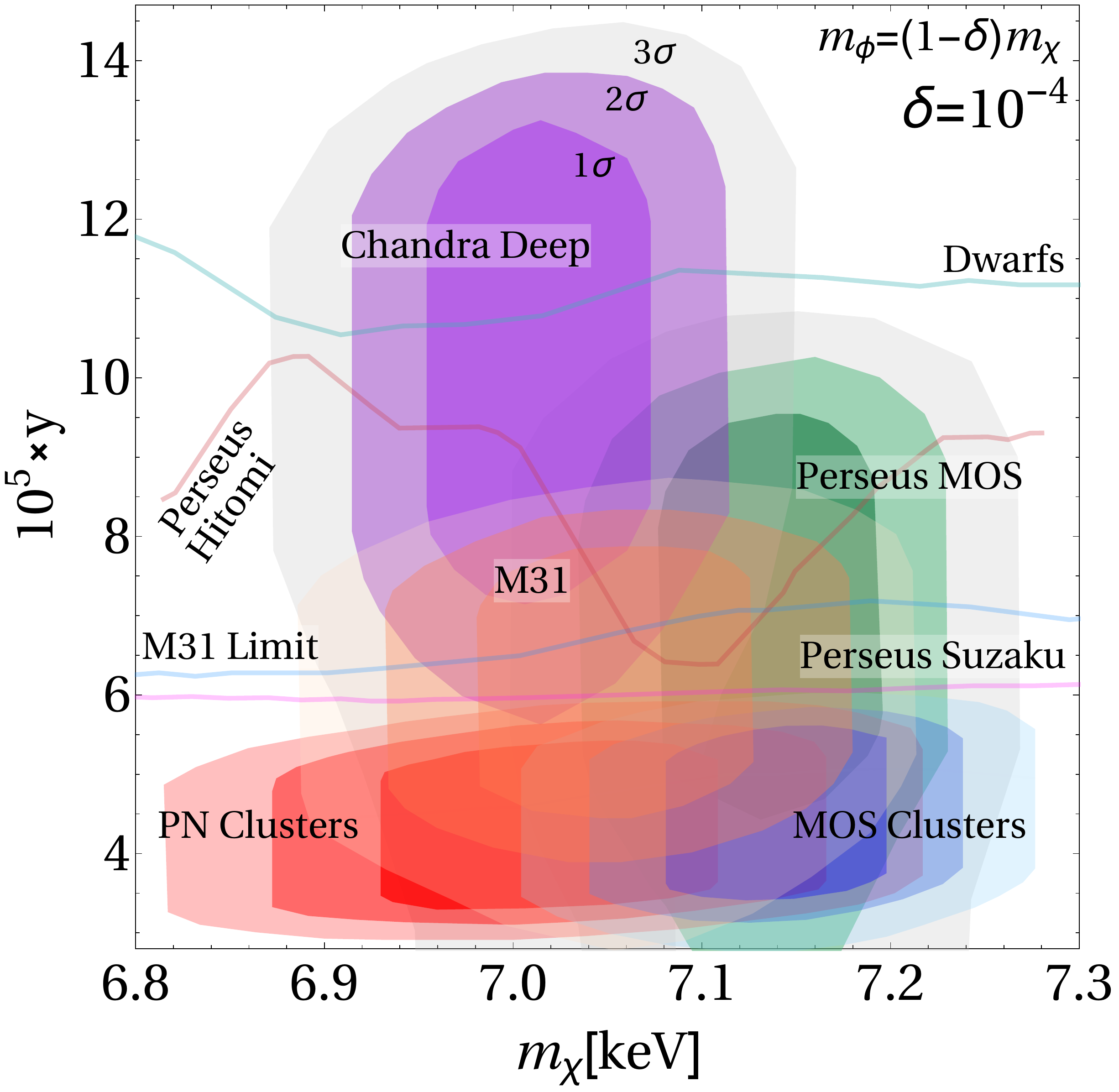}\\
        \includegraphics[width=0.4\textwidth]{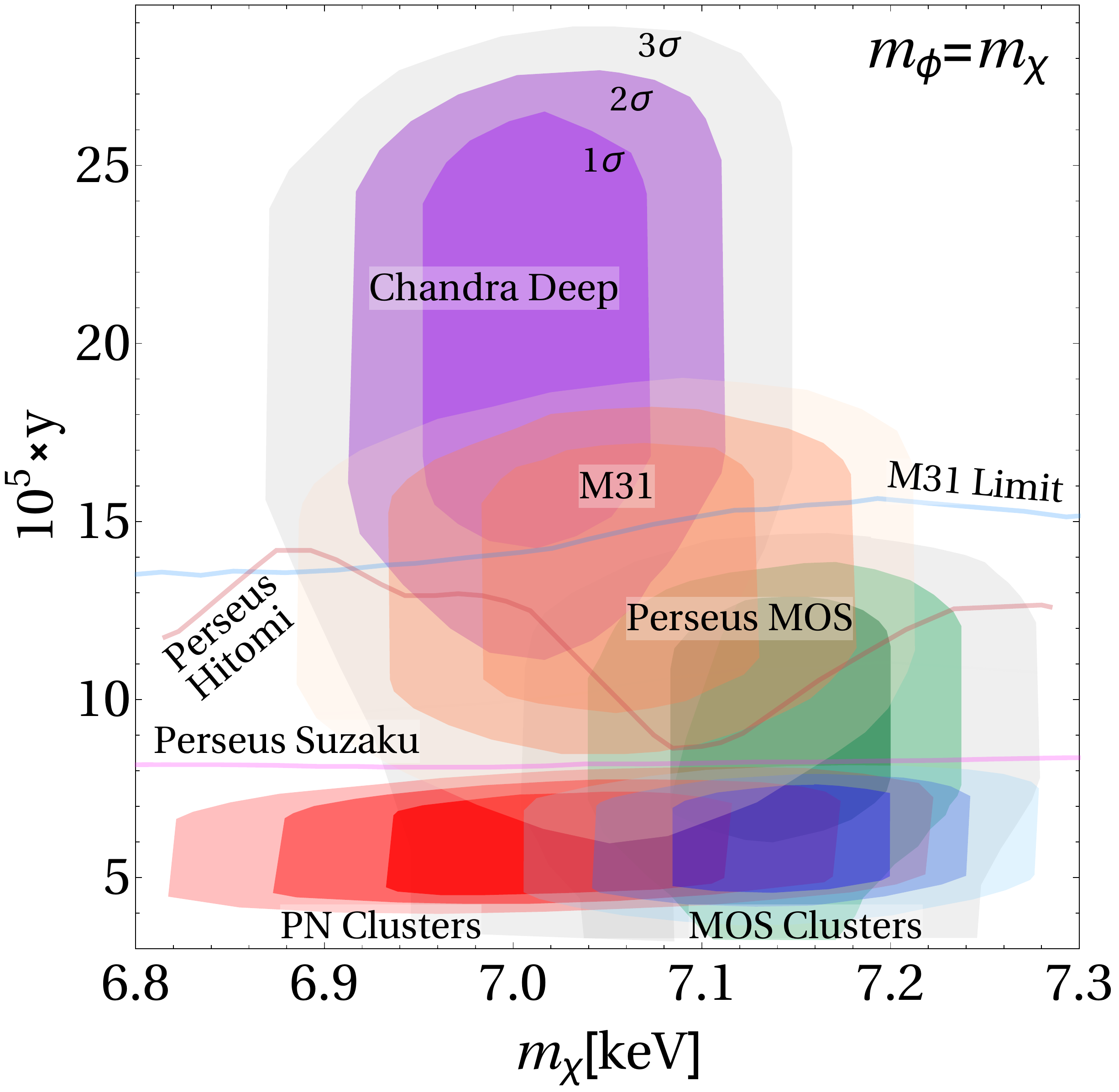}
  \end{tabular}
  \caption{Parameter regions of our annihilating DM scenario favored by
    observations of the 3.5\,keV X-ray line ($1, 2, 3\sigma$),
    and 90\% CL exclusion limits from null results. We have assumed a
    degeneracy parameter of $\delta \equiv (m_{\chi} - m_\phi) / m_{\chi} =
    10^{-4}$ in the top panel. The bottom panel is for the case that the DM
    particle $\chi$ and the intermediate scalar $\phi$ are exactly degenerate. 
    From the compilation in ref.~\cite{Abazajian:2017tcc}, we adopt
    observations from stacked galaxy clusters using the MOS and PN instruments
    on the XMM~Newton satellite~\cite{Bulbul:2014sua}, from the Perseus cluster
    alone \cite{Bulbul:2014sua}, from the Andromeda Galaxy (M31)
    \cite{Boyarsky:2014jta}, and from the Chandra Deep Field
    \cite{Cappelluti:2017ywp}. Limits are based on non-observations in M31
    \cite{Horiuchi:2013noa}, dwarf spheroidal galaxies \cite{Malyshev:2014xqa},
    and the Perseus cluster \cite{Tamura:2014mta, Aharonian:2016gzq}, and
    the regions above the solid lines are excluded.}
  \label{fig:searches}
\end{figure}

Focusing specifically on the signal at 3.5\,keV, \cref{fig:searches} demonstrates
that our scenario can align the different
observations and non-observations of the line.  We have taken into account
an uncertainty of roughly a factor of two in the DM velocity distribution for
each astrophysical target, based on varying the parameters of the underlying NFW
profiles within the ranges found in the literature~\cite{Cline:2014vsa}.
For cumulative data sets from several sources,
we sum \cref{eq:dec-ann-flux} over all of them, weighted by the individual
exposure times.  For quasidegenerate DM and $\phi$ masses with $\delta =
10^{-4}$, we find that the $1\sigma$ confidence
regions overlap for all observations except those from the Chandra Deep
Field \cite{Cappelluti:2017ywp}.  The alignment between different observations
is thus marginally better than for decaying DM~\cite{Abazajian:2017tcc}.
For $\delta = 10^{-4}$, \cref{eq:sigmav1,eq:sigmav2}
contribute roughly equally for galaxy clusters, while \cref{eq:sigmav2}
dominates for galaxies and dwarf galaxies.
When compared to exclusion limits, our scenario fares
better than decaying DM because the limit from dwarf
galaxies is less relevant thanks to the strong dependence of the annihilation
cross section on the DM velocity. We remark that future high-resolution
observations of the 3.5\,keV line~\cite{micro-x,hitomi-2} could
potentially distinguish our model from models of decaying DM and from
astrophysical explanations of the line by studying the line shape.


\vspace{1ex}
{\bf Dark Matter Production.}
In the current literature, three major mechanisms are being discussed for the
production of keV-scale DM: {\it (i)} non-resonant oscillations of active
neutrinos $\nu_a$ to sterile neutrinos $\nu_s$ (Dodelson, Widrow \cite{Dodelson:1993je}); {\it
(ii)} resonant $\nu_a \to \nu_s$ oscillations (Shi, Fuller \cite{Shi:1998km});
{\it (iii)} freeze-in of heavy particles \cite{Hall:2009bx,Klasen:2013ypa},
followed by their decay to keV-scale $\nu_s$ \cite{Merle:2013wta,
Adulpravitchai:2014xna, Konig:2016dzg}.
The parameter space for mechanisms {\it (i)} and {\it (ii)} is
heavily constrained by searches for X-ray lines  \cite{Boyarsky:2005us,
Abazajian:2006jc, Horiuchi:2013noa, Perez:2016tcq} and by structure formation
\cite{Viel:2013apy, Merle:2015vzu, Schneider:2016uqi, Menci:2017nsr, Cherry:2017dwu,
Irsic:2017ixq, Yeche:2017upn, Murgia:2017lwo} (see also \cite{Hansen:2017rxr}).

The model defined by \cref{eq:L-eff} and its UV-completion in \cref{eq:Lagrangian}
(or minimal extensions thereof, see below) enable several new DM production
mechanisms, all of them based on freeze-in of dark sector particles.
For a numerical analysis, it is convenient to parameterize the DM abundance
by the DM yield $Y \equiv n_\text{DM} / S(T)$. Here, $S(T) \equiv 2\pi^2 g^S_*(T) \, T^3 / 45$
is the entropy density of the Universe at temperature $T$,
and the number of effective relativistic degrees of freedom is
$g_*^S$~\cite{Kolb:1990}.  The Boltzmann equation governing
freeze-in for a general process $A B \to C D$
can be written as~\cite{Gondolo:1990dk,Edsjo:1997bg,Elahi:2014fsa}
\begin{align}
  \frac{dY}{dT}
    &= -\frac{1}{512 \pi^6 H(T) \, S(T)}
       \int_0^\infty \! ds\, d\Omega\, P_\text{AB} P_{C D}
                                                \nonumber \\
    &\quad\times
       \frac{\overline{|\mathcal{M}|^2}_{A B \to C D}}{\sqrt{s}}
       K_1(\sqrt{s} / T) \,.
  \label{eq:Boltzmann-2}
\end{align}
Here, $H(T) \simeq 1.66 \sqrt{g_*(T)}\,T^2 / M_\text{Pl}$
is the Hubble rate, $g_*$ is the corresponding effective number of
relativistic degrees of freedom (which for our purposes
equals $g_*^S$), and $M_\text{Pl}$ is the Planck mass. The squared
matrix element $\overline{|\mathcal{M}|^2}_{A B \to C D}$
(summed over initial and final state spins)
describes the relevant particle physics, $s$ is the center of mass energy,
the modified Bessel function of the second kind $K_1(\sqrt{s} / T)$
describes Boltzmann suppression at high $\sqrt{s}$, and
$P_\text{AB} \equiv \sqrt{[s-(m_A+m_B)^2][s-(m_A-m_B)^2] / (4s)}$ is
a kinematic factor.  If more than one $2 \to 2$ process contributes
to DM production, the right hand sides of \cref{eq:Boltzmann-2}
should be summed over all relevant processes.


\vspace{1ex}
{\bf Ultraviolet (UV) Freeze-In Through the $\phi$--Photon Couplings.} The
dominant DM production processes in the effective theory from \cref{eq:L-eff}
are $\gamma f \to \phi f$ and $\bar{f} f \to \gamma \phi$ (where $f$ denotes a
SM fermion), followed by the decay $\phi \to \bar\chi \chi$. This decay can
occur even if $m_\phi < 2 m_\chi$ because quantum corrections can raise the
effective mass of $\phi$ at high temperature.  If $\phi$ has a self-coupling of
the form $\tfrac{\lambda}{4!} \phi^4$, these corrections are of order
$(m_\phi^\text{eff})^2 \sim (\lambda/4!) (n_\phi / n_\phi^\text{eq})\, T^2$
\cite{Dolan:1973qd}, where, $n_\phi$ is the number density of $\phi$,
$n_\phi^\text{eq}$ is its value in thermal equilibrium, and $\lambda$ is a
dimensionless coupling constant.  As the $\phi$ self-coupling is not forbidden
by any symmetry, it should be included in \cref{eq:L-eff,eq:Lagrangian} anyway.

By integrating \cref{eq:Boltzmann-2} with the upper integration limit set to
the reheating temperature after inflation, $T_\text{RH}$, the DM yield is found
to be 
\begin{align}
  Y_\text{UV} \simeq
    \frac{20\,400 \, \alpha^3 \, M_\text{Pl}\, T_\text{RH}}
         {16\cdot 1.66 \cdot \pi^8 [g_*(T_\text{RH})]^{3/2} \Lambda^2}.
  \label{eq:Y-UV}
\end{align}
Here, the infrared divergence in $\gamma f \to \phi f$ has been regularized
by the effective photon mass in the plasma.  The DM abundance today is then
\begin{align}
  \Omega h^2 \big|_\text{UV}
    \simeq  105.31 \times\!
            \bigg( \frac{\text{PeV}}{\Lambda} \bigg)^2 \!
            \bigg( \frac{T_\text{RH}}{\text{TeV}} \!\bigg) \!
            \bigg( \frac{100}{g_*(T_\text{RH})} \!\bigg)^{\!\!\frac{3}{2}} \!
            \bigg( \frac{m_{\chi}}{\text{keV}} \bigg) .
  \label{eq:Oh2-UV}
\end{align}
The subscript ``UV'' in the above expressions indicates that the DM abundance
is set at $T_\text{RH}$~\cite{Elahi:2014fsa}.

The blue lines in \cref{fig:paramspace} indicate the value of $\Lambda$
needed to obtain the correct DM relic
abundance today, $\Omega h^2=0.12$~\cite{Ade:2015xua}.  We see
that a UV freeze-in scenario with $\Lambda \sim 65$\,PeV 
could explain the 3.5\,keV
line.  To the right of the dashed diagonal blue line, $\phi$ and $\chi$ come
into mutual equilibrium via $\phi \phi \leftrightarrow \bar\chi\chi$ after
freeze-in is completed. This reduces the resulting DM abundance by a factor
0.8, and slightly smaller $\Lambda$ is required to compensate. In this regime,
it is imperative that $\phi$ decays only after $\phi$ and $\chi$ have decoupled
again at $T \lesssim \text{keV}$.  Otherwise, the DM abundance would be
depleted too strongly.  We have checked that this condition is always satisfied
in the parameter regions not yet ruled out by X-ray constraints.  A potential
problem for parameter points to the right of the dashed line arises because the
photons from $\phi \to \gamma\gamma$ decays at sub-keV temperatures could leave
an observable imprint in the CMB or in the spectrum of extragalactic background
light~\cite{Cadamuro:2010cz, Cadamuro:2011fd, Cadamuro:2012rm}.  One way of
avoiding these constraints is to postulate an additional decay mode for $\phi$
that is several order of magnitude faster than $\phi \to \gamma\gamma$,
for instance a decay to pairs of light ($\ll \text{keV}$) fermions $\bar\chi' \chi'$.
In this case, the coupling $y$ in \cref{fig:paramspace,fig:searches} needs to be
rescaled by a factor $[\BR(\phi \to \gamma\gamma)]^{-1/4}$.
To avoid $\phi$ decaying while still in equilibrium with $\chi$,
$\BR(\phi \to \gamma\gamma) \gtrsim 2.1\times 10^{-3} (\text{keV}/m)^{1/3}
(y / 10^{-4} )^{16/9} (20\,\text{PeV} / \Lambda)^{10/9}$ is required,
where $y$ is the yet unrescaled coupling plotted in \cref{fig:paramspace,fig:searches}.

Note also that parameter points with $m_\phi \lesssim \text{few hundred
keV}$ and $\Lambda \lesssim 20$\,PeV
(and correspondingly lower $T_\text{RH}$ according to
\cref{eq:Oh2-UV}) are disfavored because $\phi$
production in stars could violate bounds on anomalous stellar energy
loss~\cite{Raffelt:1985nk, Raffelt:1987yu, Cadamuro:2011fd}.

Let us finally remark that in deriving \cref{eq:Y-UV,eq:Oh2-UV} we have
assumed that DM production starts only when reheating is completed and the
Universe follows standard cosmology from $T_\text{RH}$ onwards.  It is,
however, possible that reheating proceeds relatively slowly and the Universe
maintains a temperature close to $T_\text{RH}$ for a relatively long time at
the end of inflation, while Hubble expansion and inflaton decay balance each
other. In such a situation, more time is available for DM production and the
resulting $\Omega h^2$ is increased unless $\Lambda$ is increased appropriately.


\vspace{1ex}
{\bf Freeze-In Through the Misalignment Mechanism.}~\cite{Preskill:1982cy,
Nelson:2011sf, Arias:2012az}
At the end of inflation, the
field $\phi$ may be in a coherent state, with the same expectation value
$\ev{\phi}$ throughout the visible Universe. When the Hubble expansion rate
$H(T)$ drops below $m_\phi/3$, the field begins to oscillate
about its potential minimum at $\phi = 0$.  In the language of quantum field
theory, such oscillations correspond to an
abundance of $\phi$ particles.  At later times, $\phi$ and $\chi$ come into
thermal contact, and a fraction of the $\phi$ energy density is
transferred to $\chi$. ($\chi$ production via $\phi \to \bar\chi \chi$ is not
possible here since the dark sector will never become hot enough for thermal
corrections to raise $m_\phi^\text{eff}$ above $2m_\chi$.)
The resulting relic abundance of $\chi$ is
\cite{Lee:2014xua,Higaki:2014zua}
\begin{align}
  \Omega h^2 \simeq
    0.39 \times \bigg(\frac{T_{\text{RH}}}{\text{TeV}}\bigg)
                \bigg(\frac{\ev{\phi}_0}{10^{13} \,\text{GeV}}\bigg)^2 \,,
  \label{eq:misalign}
\end{align}
where $\ev{\phi}_0$ is the initial value of the field.  If the main production
channel for dark sector particles is misalignment, $\Lambda$ can be larger than
for freeze-in through the $\phi$--photon coupling.


\vspace{1ex}
{\bf Low Temperature Freeze-In.}
If the reheating temperature $T_\text{RH}$ is larger than the cutoff scale of
the effective theory, the computation of the DM abundance must be based on a UV
completion of \cref{eq:L-eff}, such as \cref{eq:Lagrangian}. In this case,
for not too small Yukawa coupling $g$, the scalar $\phi$ will be in
thermal equilibrium with $L$ and with SM particles.  $\chi$ can then
freeze in via $\phi \to \bar\chi \chi$ and $\phi\phi \to \bar\chi \chi$.
The DM abundance is approximately
\begin{align}
  \Omega h^2 &\simeq \left\{
                     \begin{array}{ll}
                       0.002 \times \bigg( \dfrac{y}{10^{-6}} \bigg)^2
                                  \bigg( \dfrac{\lambda}{10^{-8}} \bigg)^\frac{3}{2}
                                &\text{(FI via $\phi \to \bar\chi\chi$)} \\[0.4cm]
                       0.094 \times \bigg( \dfrac{y}{10^{-6}} \bigg)^4
                                &\text{(FI via $\phi\phi \leftrightarrow \bar\chi\chi$)}
                     \end{array} \right. \,.
  \label{eq:Oh2-keV}
\end{align}
The second line of \cref{eq:Oh2-keV} corresponds to the narrow green band
in \cref{fig:paramspace}.
In principle DM could also be produced directly in annihilations
of heavy vector-like leptons, $\bar{L} L \to \bar\chi\chi$.  However, for
the large $g$ and small $m_L$ required to generate the correct relic abundance
this way, $\phi$ particles would be efficiently
produced in stars (unless $m_\phi \gtrsim \text{few} \times 100$\,keV),
which is excluded ~\cite{Raffelt:1985nk,Raffelt:1987yu, Cadamuro:2011fd}.

As for UV freeze-in, $\phi$ needs to be depleted after $\chi$
production is over, for instance by introducing an invisible decay mode like
$\phi \to \bar\chi' \chi'$.  Note that $\phi$ particles abundant during Big
Bang Nucleosynthesis (BBN) at $T \sim \text{MeV}$ do \emph{not} violate the
constraint on the effective number of neutrino species, $N_\text{eff} = 2.85
\pm 0.28$~\cite{Cyburt:2015mya}.  The reason is that, by the time of BBN, the dark
sector has been sufficiently diluted by entropy production in the SM sector,
which occurs when heavy SM particles disappear from the primordial plasma.


\vspace{1ex}
{\bf Structure Formation.}
Some of the strongest constraints on keV-scale DM candidates are based
on their potential impact on structure formation as observed using
Lyman-$\alpha$ data \cite{Viel:2013apy, Merle:2014xpa, Baur:2015jsy,
  Schneider:2016uqi, Irsic:2017ixq,
Yeche:2017upn, Murgia:2017lwo, Bae:2017dpt, Bae:2017tqn} and counts of galaxies
\cite{Menci:2017nsr} and dwarf galaxies \cite{Schneider:2016uqi,
Murgia:2017lwo}.  Albeit suffering from systematic uncertainties that are
difficult to control~\cite{Kulkarni:2015fga,Cherry:2017dwu}, these observations are sensitive
to the suppression of structure at small ($\lesssim \text{Mpc}$) scales caused
by DM particles whose kinetic energy is not completely negligible yet when
structure formation begins. For DM with an initially thermal momentum spectrum,
a lower mass bound of $\gtrsim 4.65$\,keV at 95\% CL has been quoted
\cite{Yeche:2017upn}, while for sterile neutrinos produced through
oscillations, the bound is significantly stronger \cite{Yeche:2017upn,
Schneider:2016uqi} and appears to be in conflict with attempts to explain the
3.5\,keV line in such scenarios.

In our model of annihilating DM, the initial momentum spectrum of $\chi$
depends on the DM production mechanism. For freeze-in via misalignment, it
is very cold and therefore consistent with constraints. For UV freeze-in
through the $\phi$--photon coupling and for low temperature freeze-in, it is
slightly harder than thermal because the freeze-in rate is biased towards
higher energies. However, since the dark and visible sectors decouple at $T
\gtrsim 100\,\text{GeV}$, the effective temperature of the dark sector is
reduced by an entropy dilution factor $[g_*(100\,\text{GeV}) /
g_*(1\,\text{eV})]^{1/3} \simeq 2.9$ compared to the photon temperature,
improving the situation.  For $\chi$ production via $\phi \to \bar{\chi} \chi$,
each DM particles receives only half the energy of the parent particle,
reducing the DM temperature by another factor of two.  For UV freeze-in there
is moreover the possibility that the dark sector cools via self-interactions,
$\phi\phi \to 4\phi$, mediated by the same quartic coupling
$\tfrac{\lambda}{4!} \phi^4$ that enabled thermal corrections to $m_\phi$.  The
corresponding increase in $\phi$ number density can be compensated by an
appropriate adjustment of $\Lambda$, which controls the initial freeze-in. For
$m_\phi \sim 7\,\text{keV}$, choosing $\lambda \sim 10^{-3}$ boosts the
rate of $\phi\phi \to 4\phi$ above the Hubble rate at $m_\phi \lesssim T \lesssim
100$\,keV. We have checked that the inverse process $4\phi \to \phi\phi$ never
dominates over $\phi\phi \to 4\phi$, even when the dark sector turns
non-relativistic.



\vspace{1ex}
{\bf Acknowledgments.}
We are grateful to Kenny Ng for providing the astrophysical $J$-factor
underlying the NuSTAR limit in \cref{fig:paramspace}. Moreover, we have
greatly benefited from discussions with Andrew Long, Alexander Merle,
Lian-Tao Wang, Neal Weiner, and Felix Yu.
The authors have received funding from the German Research Foundation (DFG)
under Grant Nos.\ EXC-1098, \mbox{KO~4820/1--1}, FOR~2239 and GRK~1581, and from the
European Research Council (ERC) under the European Union's Horizon 2020
research and innovation programme (grant agreement No.\ 637506,
``$\nu$Directions''). JL acknowledges support by Oehme
Fellowship.


\bibliographystyle{JHEP}
\bibliography{refs}

\end{document}